\documentclass[prr,twocolumn,letterpaper]{revtex4-1}

\usepackage{amssymb,amsmath,multirow,color,bm,float,xcolor}

\usepackage[T1]{fontenc}
\usepackage{mathptmx}
\usepackage{graphicx} 

\newcommand{\pder}[2]{\frac{\partial{#1}}{\partial{#2}}}

\renewcommand{\vec}[1]{\mathbf{#1}}

\newcommand{\ez}{\vec{\hat e}_z}

\newcommand{\aave}[1]{\left\langle #1\right\rangle} 

\usepackage[english]{babel}
\makeatletter
\let\l@en\l@english
\makeatother

\renewcommand{\vec}[1]{\mathbf{#1}}

\newcommand{\ee}[1]{{\rm e}^{#1}}

\newcommand{\Va}{V_A}
\newcommand{\rms}{r.m.s.}
\newcommand{\zpm}{\vec z^\pm}

\begin{document}

\author{Jean C. Perez}
\author{Sofiane Bourouaine}
\affiliation{Florida Institute of Technology, 150 W University blvd, Melbourne, Florida, 32901, USA}

\title{The Eulerian space-time correlation of strong Magnetohydrodynamic Turbulence}

\begin{abstract}
The Eulerian space-time correlation of strong Magnetohydrodynamic (MHD) turbulence in strongly magnetized plasmas is investigated by means of direct numerical simulations of Reduced MHD turbulence and phenomenological modeling. Two new important results follow from the  simulations: 1) counter-propagating Alfv\'enic fluctuations at a each scale decorrelate in time at the same rate in both balanced and imbalanced turbulence; and 2) the scaling with wavenumber of the decorrelation rate is consistent with pure hydrodynamic sweeping of small-scale structures by the fluctuating velocity of the energy-containing scales. An explanation of the simulation results is proposed in the context of a recent phenomenological MHD model introduced by Bourouaine and Perez 2019 (BP19) when restricted to the strong turbulence regime. The model predicts that the two-time power spectrum exhibits an universal, self-similar behavior that is solely determined by the probability distribution function of random velocities in the energy-containing range. Understanding the scale-dependent temporal evolution of the space-time turbulence correlation as well as its associated universal properties is essential in the analysis and interpretation of spacecraft observations, such as the recently launched \emph{Parker Solar Probe (PSP)}.

\end{abstract}

\maketitle

\section{Introduction}

Since the first observation that non-compressive Alfv\'en-like fluctuations of velocity and magnetic field dominate the solar wind by~\citet{belcher71}, incompressible Magnetohydrodynamics (MHD)~\cite{biskamp03} has been often invoked to describe the observed Kolmogorov-like the power spectrum of low-frequency fluctuations of the solar wind plasma, for an extensive review see~\citet{bruno13,chen16,verscharen19}.  
The majority of advances in MHD turbulence in the last few decades have been largely concerned with its spatial statistical properties, such as the three dimensional  structure of the power spectrum and higher order structure functions~\citep{biskamp03,bruno13,chen16,verscharen19,galtier00,muller05,lithwick07,boldyrev05,mason08,perez08,perez09,perez12,mason12,chandran15,mallet17a,boldyrev17,loureiro17,beresnyak14}. Most of these properties can be derived from two-point one-time correlations, which quantify the covariance between simultaneous values of a turbulent quantity at two different points.

However, more often than not turbulence experiments and solar wind observations can only provide single-probe measurements along the plasma at different times and locations, in which case a methodology to relate the time signals measured in the probe-frame to the spatial properties in the plasma-frame is required to test theoretical predictions. For instance, in solar wind observations the so called Taylor Hypothesis (TH)~\cite{taylor38} (or frozen-in-flow approximation) is commonly used. This approximation essentially establishes that when the mean flow speed $U$ (as seen in the probe frame) is much larger than any other characteristic speed, such as the flow's turbulent amplitude $u_0$ and wave-propagation speed, the time signal of a turbulent quantity measured in the probe's frame is due to the advection of a frozen spatial structure passing by the instrument at the local flow speed $U$. In contrast, when these conditions are violated the temporal variation observed single-probe measurements arise instead from a \emph{dynamic} structure passing by the probe, i.e., the time variation is a combination of advection and evolution of the passing structures. The recently launched \emph{Parker Solar Probe} (\emph{PSP})~\citep{fox16} has spurred a renewed interest in understanding the space-time structure of solar wind turbulence~\cite{narita17,bourouaine18,bourouaine19,chhiber19,bourouaine20}, precisely because it will explore the near-Sun region where the conditions for the validity of the TH might not be satisfied. In this case, an understanding of the structure of two-time two-point correlations of turbulent quantities and any possible universal properties are essential to successfully relate the turbulent time signals (measured by the probe) to the spatial structure of the turbulence.  Analyses of spacecraft data to date have provided increasing evidence that many turbulent properties of low frequency fluctuations in the solar wind are consistent with various predictions of current MHD turbulence models~\cite{chen16}, however the subject remains open and under active investigation and debate. In this paper we address the problem of the physics of temporal decorrelation of Alfv\'enic fluctuations, which may be relevant to the analysis of  solar wind fluctuations whenever they can be described by incompressible MHD~\cite{narita17,bourouaine18,bourouaine19,chhiber19,bourouaine20,servidio11,lugones16}, such as in the first two perihelia around $36$ solar radii where the solar wind was found to be highly Alfv\'enic~\cite{bale19,kasper19,chen20,dudokdewit20,parashar20,tenerani20,mcmanus20}. The physics of the temporal decorrelation in solar wind observations when MHD is not applicable is outside the scope of this work and deserves further investigation.


A number of works on the structure of the Eulerian space-time correlation in MHD turbulence have been carried out in the past decade. \citet{servidio11} investigated the scale-dependent temporal correlation in isotropic MHD using numerical simulations. The authors reported that the Eulerian decorrelation time is consistent with a sweeping-like scaling $\tau_c\sim k^{-1}$, which they attribute to a combination of convective sweeping by the large-scale flow and magnetic sweeping from large-scale magnetic fluctuations.  \citet{lugones16} studied the same problem for the anisotropic case of MHD turbulence with a guide field, using simulations with small, moderate and large guide field to investigate the role of the magnetic field in the decorrelation time. Their findings are consistent with~\citet{servidio11} for small fields, while the decorrelation becomes dominated by Alfv\'en-wave propagation for large guide field. 
One shortcoming of the works of~\citet{servidio11} and \citet{lugones16} is that they focused on the correlation function of the fluctuating magnetic field and not on the Elsasser fields. \citet{narita17a} investigated the temporal decorrelations of the Elsasser fields in MHD, by extending a Hydrodynamic (HD) sweeping model of the Eulerian correlation of~\citet{wilczek12} with a mean flow. His model suggests that the temporal decorrelations of the Elsasser fields $\vec z^\pm$ at small-scales arises from the random sweeping by large-scale Elsasser fluctuations propagating in the opposite direction $\vec z^\mp$, resulting in different decorrelation rates for imbalanced turbulence. However, \citet{bourouaine18} measured the Eulerian correlation of Elsasser fields in highly imbalanced, reflection-driven MHD turbulence simulations with high space-time resolution and found that the decorrelation of both fields is consistent with sweeping by large-scale fluctuations at a common speed that is comparable to the root mean squared (\rms) value of the fluctuating velocity, suggesting that the sweeping is hydrodynamic in nature. Based on the evidence from the numerical simulations, \citet[][hereafter BP19]{bourouaine19} recently introduced a new sweeping model of MHD turbulence that relies on the local mean field direction. The findings from the model are consistent with a common sweeping characteristic timescale for both Elsasser fields. In this work we show that when the BP19 model is applied in strong MHD turbulence, the Eulerian space-time correlation is entirely dominated by HD sweeping, which we confirm in numerical simulations. 

This paper is organized as follows. In section~\ref{sec:theory} we discuss the theoretical framework for the simulations and phenomenological models presented in this work, including a brief description of Kraichnan's idealized convection model in HD and its generalization to strong MHD turbulence. In section~\ref{sec:numerics} we discuss the numerical simulation setup and simulation parameters and a brief description of the methodology that will be used to validate the MHD sweeping model in simulations. In section~\ref{sec:results} we present and discuss the simulation results and in section~\ref{sec:conclusion} we conclude.  

\section{Theoretical Framework\label{sec:theory}}
 We assume that velocity $\vec v(\vec x,t)$ and magnetic field $\vec B(\vec x,t)$ fluctuations are described by the equations of ideal incompressible MHD, which in terms of the Elsasser variables $\vec z^\pm\equiv\vec v\pm\vec b$ take the form
 \begin{equation}
     \left(\pder{}t\mp\vec\Va\cdot\nabla\right)\vec z^\pm=-\vec z^\mp\cdot\nabla\vec z^\pm-\frac 1\rho\nabla p+{\vec{ f}}^\pm+\nu\nabla_\perp^2\vec{z}^\pm,\label{eq:MHD_elsasser}
 \end{equation}
where $\vec b=\vec B/(4\pi\rho)$ is the fluctuating Alfv\'en velocity, $\vec\Va={\bf B}_0/(4\pi\rho)$ is the background Alfv\'en velocity, $\rho$ is the constant background plasma density and $p$ is the combined thermal and magnetic pressure. Random forcing $\vec f^\pm$ and viscous dissipation terms have been included to investigate the case of steadily-driven turbulence. 

For a strong background magnetic field ($\vec B_0=| \vec B_0|\ez$, say, with $|\vec B_0| \gg|\vec b|$) the universal properties of MHD turbulence can be accurately described by neglecting the field-parallel component, $\vec z^\pm_\|$, of the fluctuating fields (the pseudo-Alfv\'en fluctuations) that play a sub-dominant role in the turbulence dynamics (see~\cite{perez12} and references therein).
It can be further demonstrated that setting $\vec z_\|=0$ in equation~\eqref{eq:MHD_elsasser} leads to a set of equations that is equivalent the simpler Reduced MHD (RMHD) model~\cite{kadomtsev74,strauss76}. It is worth noting that the RMHD model is commonly invoked to describe the dominant nonlinear interactions and resulting turbulence of non-compressive Alfv\'en-like fluctuations, which comprise most of the energy in the solar wind. It has also been shown, from gyrokinetics~\cite{schekochihin09} and from comparisons with MHD simulations~\cite{mason11,mason12}, that RMHD rigorously describes the essential non-linear interactions responsible for the turbulence cascade of non-compressive Alfv\'enic fluctuations.

For homomogeneous and stationary Elsasser fluctuations $\zpm(\vec x,t)$ the two-time two-point Eulerian correlation is defined as
\begin{equation}
    C^\pm(\vec r,\tau)\equiv\aave{\zpm(\vec x,t)\cdot\zpm(\vec x+\vec r,t+\tau)},\label{eq:correlation}
\end{equation}
where $\aave{\cdots}$ represents an ensemble average over many turbulence realizations. These correlations measure the degree to which each Elsasser field at any position $\vec x$ and time $t$ is correlated with itself at another location with relative position $\vec r$ after a time $\tau$ has elapsed.  In a turbulent system correlations arise due to the presence of coherent structures of many characteristic length-scales, which are undergoing random advection and nonlinear straining by other structures in the flow according to the dynamics determined by equations~\eqref{eq:MHD_elsasser}. Although turbulence correlations can, at least formally, be related to the governing equations of the fluctuating variables, turbulence theories have been unable to produce exact analytical solutions even in the simplest case of incompressible HD turbulence, because of the well known closure problem~\cite{frisch95}.

The correlation in equation~\eqref{eq:correlation} can be expressed in terms of its spatial Fourier transform
\begin{equation}
  C^\pm(\vec r,\tau)=\int h^\pm(\vec k,\tau)\ee{i\vec k\cdot\vec r}d^3k,\label{eq:correlation2}
\end{equation}
where $h^\pm(\vec k,\tau)$ are the so-called two-time power spectra
\begin{equation}
    h^\pm(\vec k,\tau)=\aave{\zpm(-\vec k,t)\cdot\vec \zpm(\vec k,t+\tau)},\label{eq:ktspectrum}
\end{equation}
and $\zpm(\vec k,t)$ is the spatial Fourier transform of the field $\zpm(\vec x,t)$. 
The scaling properties and three-dimensional structure of the spatial power spectra $h^\pm_0(\vec k)=h^\pm(\vec k,0)$ (for $\tau=0$) have been the subject of extensive research in theory, numerical simulations and solar wind observations~\citep{chen16}. In this work, we make very few assumptions about the structure of the spatial power spectrum and focus our investigation on the structure of the scale-by-scale $\tau$ dependency, which accounts for the scale-dependent temporal decorrelation of the turbulence. As $\tau$ increases, the Fourier amplitudes at wavevector $\vec k$ decorrelate and one can thus define the \emph{scale-dependent time correlations} $\Gamma^\pm(\vec k,\tau)$ as
\begin{equation}
 h^\pm(\vec k,\tau)=h^\pm_0(\vec k)\Gamma^\pm(\vec k,\tau).\label{eq:hgamma}
\end{equation}
By definition $\Gamma^\pm(\vec k,0)=1$ for all $\vec k$, which means that at zero time lag $\tau$ the fluctuations are perfectly correlated.  The advantage of equation~\eqref{eq:hgamma} is that it allows for the separation of the spatial part of the correlation function from the scale-dependent temporal part, which is the subject of this work. The two-time power spectra $h^\pm(\vec k,\tau)$ are simply a different representation of the correlation functions $C^\pm(\vec r,\tau)$ and thus contain the same information.   

\subsection{Kraichnan's idealized convection model in Hydrodynamics}
In HD turbulence, temporal decorrelation at a given point can arise from two main effects: 1) random sweeping of the small-scale eddies by large ones and 2) eddy straining (or shear) associated with nonlinear inertial forces.  Scaling arguments can be used to argue that the Eulerian correlation in Hydrodynamic (HD) is dominated by the first of these two effects, also known as the Kraichnan's Sweeping Hypothesis (KSH)~\cite{kraichnan64}. The sweeping decorrelation mechanism is a non-local-in-scale process, in the sense that it involves eddies of disparate scales, and its characteristic timescale is $\tau_{\rm s}\sim 1/(ku_0)$ for a fluctuation of scale $\lambda\sim 1/k$ swept by a large-scale fluctuation with velocity $u_0$. The second timescale associated with nonlinear straining scales as $\tau_{NL}\sim k^{-2/3}$, a slower decrease with $k$ than the sweeping timescale $\tau_{\rm s}$. These scaling arguments suggest that the Kraichnan's hypothesis is expected to hold better for sufficiently large values of $k$ for which the ratio $\tau_{\rm s}/\tau_{\rm NL}\sim k^{-1/3}$ is small. 
   
Kraichnan introduced an idealized convection model of incompressible HD to describe the random sweeping of small-scale fluctuations by large ones. In this model the fluid velocity consists of two parts $\vec v=\vec v'+\vec u$, with the following assumptions: 1) $\vec v'$, describing the large-scale eddies, is constant in space and time but is a zero-mean random variable with an isotropic Gaussian distribution, 2) the field $\vec u(\vec x,t)$, describing the small-scale eddies, is much smaller in magnitude than $\vec v'$, and 3) $\vec v'$ and $\vec u(\vec x,0)$ are statistically independent. From the first two assumptions, the Navier-Stokes equation becomes
\begin{equation}
    \pder{\vec u}t+(\vec v'+\vec u)\cdot\nabla\vec u\simeq \left(\pder{}t+\vec v'\cdot\nabla\right)\vec u= 0,\label{eq:HD2}
\end{equation}
where we ignored viscous dissipation (considering fluctuations in the inertial range) and dropped the pressure term whose only role is to ensure fluctuations remain incompressible. As opposed to the Navier-Stokes (NS) equation, the idealized model given by equation~\eqref{eq:HD2} is a stochastic linear equation in $\vec u$, which does not have the statistical closure problem of the nonlinear NS equation. The essence of this approximated idealized model is that the dominant variation of $\vec u$ simply arises from advection of frozen structures by a constant but random velocity at each point. For instance, if we assume for the moment that $\vec v'=\vec V$ is not a random variable, equation~\eqref{eq:HD2} forms the basis for the TH approximation.
In this sense, the random sweeping model can be interpreted as the application of TH to a statistical ensemble of systems, each one with a different large scale flow velocity drawn from a random distribution corresponding to the large-scale eddies. Straightforward solutions to Equation~\eqref{eq:HD2} can be found to obtain the scale-dependent time correlation
\begin{equation}
    \Gamma(\vec k,\tau) = \aave{\ee{-i\vec k\cdot\vec v'\tau}}=\int \ee{-i\vec k\cdot\vec v'\tau}P(\vec v')d^3v',\label{eq:gammakt}
\end{equation}
where $P(\vec v')$ is the probability density for the random variable $\vec v'$. Kraichnan's model assumed $P(\vec v')$ to be an isotropic Gaussian distribution
\begin{equation}
        P(\vec v')=\frac 1{(2\pi v_0^2)^{3/2}}\exp{\left(-\frac{|\vec v'|^2}{2v_0^2}\right)},\label{eq:Pgauss}
\end{equation}
in which case equation~\eqref{eq:gammakt} becomes
\begin{equation}
    \Gamma(\vec k,\tau) = \ee{-\gamma_k^2\tau^2},
    \label{eq:gamma}
\end{equation}
where $\gamma_k\equiv kv_0/\sqrt 2$ is the decorrelation rate and $v_0$ is the \rms~value of the velocity $\vec v'$ along any given direction.  The decorrelation rate is defined at each $\vec k$ as $\gamma_k=1/\tau_k$, where $\tau_k$ is the time lag for which the correlation $\Gamma(\vec k,\tau)$ drops to $1/e\simeq 0.37$. This idealized model provides a phenomenological description of the temporal decorrelation when the timescale $\tau_c\sim 1/(kv_0)$ is much faster than the Kolmogorov estimate of the nonlinear cascade time $\tau_{NL}\sim k^{-2/3}$. \citet{wilczek12} revisited the HD case with a constant mean flow $\vec U$, which simply adds a  phase factor to the correlation $\Gamma(\vec k,\tau)$.
    
It is important to note that Kraichnan's assumption of Gaussianity for the random variable $\vec v'$ is not necessary and the validity of his model can be extended to other distributions $P(\vec v')$  by noticing that the average in equation~\eqref{eq:gammakt} is nothing but the characteristic function $\varphi_{\vec v'}(\boldsymbol\xi)$ of the probability density $P(\vec v')$, hence
\begin{equation}
    \Gamma(\vec k,\tau)=\varphi_{\vec v'}(\vec k\tau),
\end{equation}
where
\begin{equation}
    \varphi_{\vec v'}(\boldsymbol\xi)\equiv\aave{\ee{-i\boldsymbol\xi\cdot\vec v'}}=\int\ee{-i\boldsymbol\xi\cdot\vec v'}P(\vec v')d^3v'.\label{eq:pchar}
\end{equation}

This result shows that the scale-dependent time correlation can be obtained from characteristic function of the probability density of the large-scale eddies~\citep{lumley65}, by setting the velocity-wavenumber $\boldsymbol\xi$ equal to $\vec k\tau$, and is therefore self-similar. In the next subsection we extend this idealized model for the case of strong MHD turbulence following the phenomenology introduced by BP19.

\subsection{Sweeping model for strong MHD turbulence.} 

\begin{table*}[!ht]
  \begin{center}
    \begin{tabular}{ccccccccccc}
      \hline\hline
      Run  & Regime     & Normalized cross-helicity ($\sigma_c$) & $u_0$ & $b_0$ & $z^+_0$& $z^-_0$ & Aspect ratio & Resolution & Re \\
      \hline\hline
      RB1  & Balanced   & 0.2   & 0.81  & 1.06  & 1.22 & 1.43 & 1:6  &$512^3$ &2400 \\
      RB2  & Balanced   & 0.0   & 0.79  & 1.12  & 1.36 & 1.37 & 1:6  &$1024^3$ &6000 \\
      RI1  & Imbalanced & 0.5   & 0.74  & 0.97  & 1.48 & 0.9  & 1:10 &$512^3$ & 2400\\
      
      \hline
    \end{tabular}
    \caption{Simulation list and relevant parameters as follows: normalized cross-helicity $\sigma_c$, \rms values of $\vec v,\vec b,\vec z^\pm$ of most energetic scales, aspect ratio $1:M$, numerical resolution and Reynolds number.\label{tab:sims}}
  \end{center}
\end{table*}

The Kraichnan's picture acquires greater complexity in MHD turbulence for a number of reasons. First, in the Elsasser formulation MHD contains two fluctuating fields $\vec z^\pm$ that are being advected in opposite directions along the background magnetic field and undergo mutual straining only when counter-propagating fields encounter each other or ``collide'', resulting in various limiting regimes. For instance, when the energies of the fluctuating fields $\vec z^+$ and $\vec z^-$ are comparable  the turbulence is called \emph{balanced}, otherwise it is called \emph{imbalanced}. For both the balanced and imbalanced cases, the turbulence can be weak~\citep{galtier00} or strong~\citep{goldreich95}. The weak turbulence regime occurs when the time it takes two eddies to cross one another is much shorter than the nonlinear interaction time, thereby requiring a large number of successive collisions before eddies can cascade their energy to smaller scales~\cite{galtier00}. In the strong regime, the crossing and nonlinear times are comparable and the cascade occurs in a single collision. Although it is still a matter of debate, a number of models assume that for imbalanced turbulence $\vec z^\pm$ may have different nonlinear straining times $\tau^\pm_{\rm NL}\sim \lambda/z^\mp_\lambda$, due to the amplitude difference~\citep{lithwick07,beresnyak08,chandran08}. 

BP19 introduced a new model for the scale-dependent time correlations, by writing the large-scale Elsasser variables ${\vec z'}^\pm=\vec v'\pm\vec b'$ in terms of their corresponding fluctuating velocity and magnetic fields in~\eqref{eq:MHD_elsasser} to obtain
\begin{equation}
     \left(\pder{}t\mp\vec\Va'\cdot\nabla_\|+{\vec v'}\cdot\nabla_\perp\right)\delta\vec z^\pm=-\delta\vec z^\mp\cdot\nabla_\perp\delta\vec z^\pm,\label{eq:MHD_elsasser3}
\end{equation}
where $\vec v',\vec b'$ and $\delta\vec z^\pm$ are taken to be perpendicular to the magnetic field, consistent with the RMHD approximation. Here $\delta\zpm$ represent the small-scale Elsasser fluctuations, $\vec\Va'\equiv\vec\Va+\vec b'$ 
is the modified Alfv\'en velocity resulting from the superposition of the mean background magnetic field and the fluctuating component $\vec b'$ from the large-scale eddies, and $\nabla_\perp,\nabla_\|$ are the field-perpendicular and field-parallel gradient operator defined with respect to the local magnetic field, respectively. Hereafter, primes are used to represent random variables with known statistics.  
Under the assumption that the characteristic timescales of the RHS terms are much smaller than those in the LHS, for a strongly magnetized plasma ($|\vec b'|\ll\Va$) and for Gaussian-distributed outer-scale velocities $\vec v'$ one obtains~\citep{bourouaine19}
\begin{equation}
    \Gamma^\pm(\vec k,\tau)=\ee{\mp ik_\|\Va\tau}\ee{-(\gamma_k\tau)^2}\label{eq:bp19}
\end{equation}
where $\Va$ is the Alfv\'en speed, $k_\|$ is the component of $\vec k$ in the direction of the local magnetic field and $\gamma_k=k_\perp v_0/\sqrt 2$ is the decorrelation rate. This result is very similar to the model obtained by~\citet{narita17a}, with the important difference that both Elsasser fields decorrelate at a common rate, determined by pure HD sweeping. Noting that $v_0$ represents the \rms~of the fluctuating velocity in any direction and $\vec v'$ lies in the field-perpendicular plane, the velocity~\rms~is $u_0=\sqrt 2v_0$ in which case $\gamma_k=k_\perp u_0/2$.

Scaling arguments can also be used to obtain a model for the $\Gamma^\pm(\vec k,\tau)$ functions in the strong turbulence regime.  
Let $\lambda$ and $l$ be the field-perpendicular and field-parallel lengthscales of an eddy with respect to the local magnetic field and of amplitude $v_\lambda$. 
If the turbulence is driven isotropically, $\lambda\sim l$, the turbulence is necessarily weak when $v_{\lambda}\ll\Va$. Because weak turbulence cascades energy to smaller perpendicular scales without affecting the parallel structure $l$~\citep{galtier00}, eddies will progressively become elongated along the field until the nonlinear time $\tau_\lambda\sim\lambda/v_{\lambda}$ becomes comparable to the linear timescale $\tau_A\sim l/\Va$. Therefore, the turbulence will unavoidably become strong when the \emph{critical balance} condition~\cite{goldreich95} $\tau_{\lambda}\sim\tau_A\Rightarrow\lambda/v_{\lambda}\sim l/\Va$ is satisfied, which means the timescale of Alfv\'en wave propagation becomes comparable to that of the nonlinear terms in the RHS of equation~\eqref{eq:MHD_elsasser3}. In this case, the Alfv\'enic propagation can be neglected and the correlation function takes the form
\begin{equation}
    \label{eq:two-time-spectrum-char}
    \Gamma^\pm(\vec k,\tau)=\varphi_{\vec v'}(\vec k_\perp\tau),
\end{equation}
and in the case of a Gaussian distribution $P(\vec v')$
\begin{equation}
    \Gamma^\pm(\vec k,\tau)=\ee{-(\gamma_k\tau)^2}
    \label{eq:two-time-spectrum}.
\end{equation}
Equations~\eqref{eq:two-time-spectrum-char}~and~\eqref{eq:two-time-spectrum} show that the decorrelation is solely determined by HD sweeping and is the same for both Elsasser fields.

It is worth mentioning here that the assumption of a Gaussian distribution of large-scale velocities is made here for concreteness, to make calculations simpler and because the simulations used in next section to validate the model are driven at the outer-scale in a Gaussian fashion. However, for the application of this model to solar wind observations~\citep[see for instance][]{bourouaine20}, the actual distribution of velocities at the outer-scale can be used. Previous solar wind observations have shown evidence that large-scale velocities and magnetic field fluctuations are Gaussian~\cite{bruno13}, although some observations suggest these fluctuations can show strongly non-Gaussian, skewed tails~\cite{marino12}.

\section{Numerical Simulations\label{sec:numerics}} 
RMHD equations, obtained by setting $\vec z_\|=0$ in Eq.~\eqref{eq:MHD_elsasser},  are solved using a fully dealiased 3D pseudo-spectral code in a rectangular domain with aspect ratio $1:M$, where $M$ is the ratio between parallel and perpendicular box sizes, defined with respect to the background magnetic field $\vec B_0=B_0\ez$.  The normalization chosen in the simulations is such that the \rms~values of fluctuating plasma and Alfv\'en velocity are of order $u_{\rm rms}\sim 1$ (in code's units), and the magnitude $B_0$ so that $\Va/u_{\rm rms}$ is of order $M$.  The box size in the $xy$ plane (perpendicular to the guide magnetic field) is chosen as $L = 2\pi$ and time is normalized to the large scale eddy turnover time $\tau_0=L/2\pi u_{\rm rms}$. The turbulence is driven by random forcing in the field-perpendicular and field-parallel wave-numbers $0<k_\perp<4$ and $0<k_\|<2$, respectively, which due to the aspect ratio of the simulation box allows one to drive strong RMHD turbulence by controlling the degree to which outer-scale eddies satisfy the critical balance condition $k_\|\Va\sim k_\perp u_{\rm rms}$~\cite{perez10}, where $u_{\rm rms }$ represents the \rms~value of the turbulent velocity. The Reynolds number is defined as ${\rm Re}=u_{\rm rms}(L/2\pi)/\nu$. Approximately every eddy turnover time in the steady state the code outputs snapshots of the Elsasser fields $\vec z^\pm$, as well as the entire time history of each Elsasser field on eight selected $xy$ planes. In the simulations, correlations between $\vec v$ and $\vec b$ are introduced through the random forcing to investigate the role of cross-helicity $H_c=E^+-E^-$, which measures the energy difference between counter-propagating Elsasser fluctuations. Cross-helicity is conveniently quantified in the simulations through the normalized cross-helicity $\sigma_c=H_c/E$ defined as the amount of cross-helicity normalized to the total energy $E=E^++E^-$. The normalized cross-helicity takes values in the range $-1\le\sigma_c\le 1$, with zero corresponding to \emph{balanced} turbulence, and \emph{imbalanced} turbulence otherwise.

Three simulations of steadily-driven RMHD turbulence, listed in table~\ref{tab:sims}, are used to investigate the scaling properties of the time  correlations $\Gamma^\pm(\vec k,\tau)$ and to compare with phenomenological models . These simulations have been extensively used to investigate the structure and scaling of the spatial spectrum $h_0(\vec k)$ of balanced and imbalanced MHD turbulence in previous works~\cite{perez08,perez09,perez12}. Because the parameters of the simulations in table~\ref{tab:sims} are the same as simulations RB1, RB2 and RI1 in~\citet{perez12}, we adopt the same labeling convention. 

\begin{figure}[!t]
  \centering
  \includegraphics{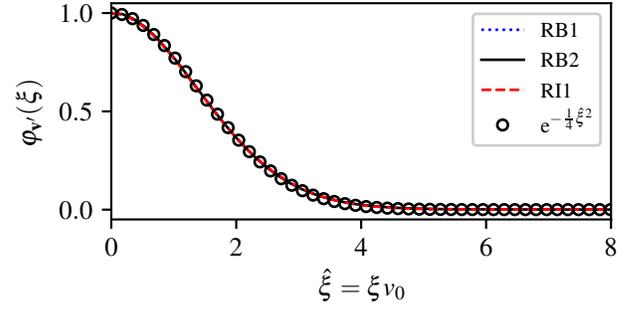}
  \caption{Angle-averaged characteristic function $\varphi_{\vec v'}$ vs normalized velocity-wavenumber $\hat\xi\equiv\xi u_0$, associated with the random distribution of large-scale velocities $\vec v'$ in the plane perpendicular to the background magnetic field ($xy$--plane). The statistical nature of the eddies in the driving range is the same in all three simulations, and only differ by the \rms~values of $u_0$ listed in table~\ref{tab:sims}. }
  \label{fig:pchar}
\end{figure}

The random forcing drives the outer-scale velocities toward an isotropic two-dimensional Gaussian distribution of the form given in equation~\eqref{eq:Pgauss}, and whose characteristic function is
\begin{equation}
    \varphi_{\vec v'}(\boldsymbol\xi)=\ee{-\frac 14u_0^2|\boldsymbol\xi|^2},
\end{equation}
where $u_0$ is the \rms~value of velocities in the energy-containing range and $\boldsymbol\xi$ is a velocity-wavevector on which the characteristic function depends. The Gaussian nature of the outer-scale flow is verified by measuring the angle-averaged characteristic function $\varphi_{\vec v'}(\xi)=\aave{\varphi_{\vec v'}(\boldsymbol\xi)}_\phi$ for the most energetic fluctuations, taken as wavenumbers $k_\perp\lesssim 7$. Figure~\ref{fig:pchar} shows that when the characteristic functions $\varphi_{\vec v'}$ for all three simulations are plotted versus $\hat\xi\equiv\xi u_0$, they all overlap almost exactly with the Gaussian function
\begin{equation}
    g(\hat\xi) = \ee{-\frac 14\hat\xi^2},
\end{equation}
represented by the circles in the figure.

The two-dimensional, two-time power spectrum of each Elsasser field
  \begin{equation}
      h^\pm_{\rm 2D}(\vec k_\perp,\tau)=\int_{-\infty}^\infty h(\vec k_\perp,k_\|)dk_\|\label{eq:h2d}
  \end{equation}
is calculated through the average
 \begin{equation}
     h_{\rm 2D}^\pm(k_\perp,\tau)=\aave{\vec z_\alpha(-\vec k_\perp,t)\cdot\vec z_\alpha(\vec k_\perp,t+\tau)}_\phi
 \end{equation}
 in terms of the Fourier transforms $\zpm_\alpha\equiv\vec z^\pm(\vec k_\perp,\zeta_\alpha,t)$ at each transverse plane $\zeta_\alpha$, with $\alpha=1,\cdots 8$. The ensemble average $\aave{\cdots}_\phi$ in this equation also includes an average over the polar angle in the $\vec k_\perp$ plane, due to the isotropy of the two-dimensional power in the field-perpendicular plane. 
 \begin{figure}[!t]
  \centering
  \includegraphics{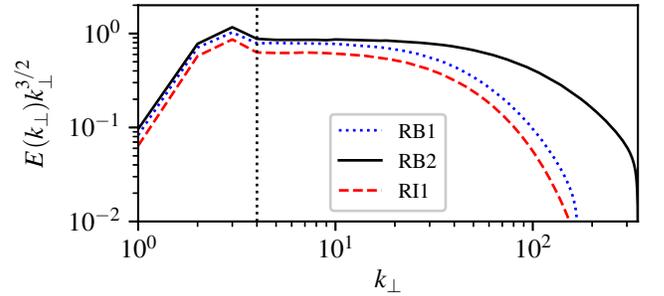}
  \caption{
  Field-perpendicular energy spectra compensated by $k_\perp^{3/2}$ for the three simulations. All compensated spectra become flat at $k_\perp\simeq 4$, consistent with power-law scaling $\propto k_\perp^{-3/2}$ for $k_\perp\gtrsim 4$. The vertical dotted line indicates the beginning of the inertial range.}
  \label{fig:h2d}
\end{figure}
The scale-dependent time correlations can be related to the two-dimensional two-time power spectra by integrating equation~\eqref{eq:hgamma} over $k_\|$
\begin{equation}
    h^\pm_{\rm 2D}(k_\perp,\tau)=\int_{-\infty}^\infty h_0^\pm(k_\perp,k_\|)\Gamma^\pm(k_\perp,k_\|,\tau)dk_\|.
\end{equation}
Assuming that the scale-dependent time correlations $\Gamma^\pm(k_\perp,k_\|,\tau)$ weakly depend on $k_\|$ as predicted by the model it follows that
\begin{equation}
    \Gamma^\pm(k_\perp,\tau) = \frac{h^\pm_{\rm 2D}(k_\perp,\tau)}{P^\pm(k_\perp)}\label{eq:g2d}
\end{equation}
where
\begin{equation}
    P^\pm(k_\perp)=\int_{-\infty}^\infty h_0^\pm(k_\perp,k_\|)dk_\|\label{eq:2dpower}
\end{equation}
is the  two-dimensional power spectrum. The field-perpendicular energy spectra, defined as
\begin{equation}
    E^\pm(k_\perp)=2\pi k_\perp P^\pm(k_\perp)
\end{equation}
measured for simulations RB1, RB2 and RI1 are shown in Figure~\ref{fig:h2d}. As previously reported, the simulations are consistent with the scale-dependent phenomenology of strong MHD turbulence~\cite{boldyrev05} for balanced and imbalanced turbulence~\cite{perez09}.

\begin{figure}[!ht]
  \centering
  \includegraphics[width=0.5\textwidth]{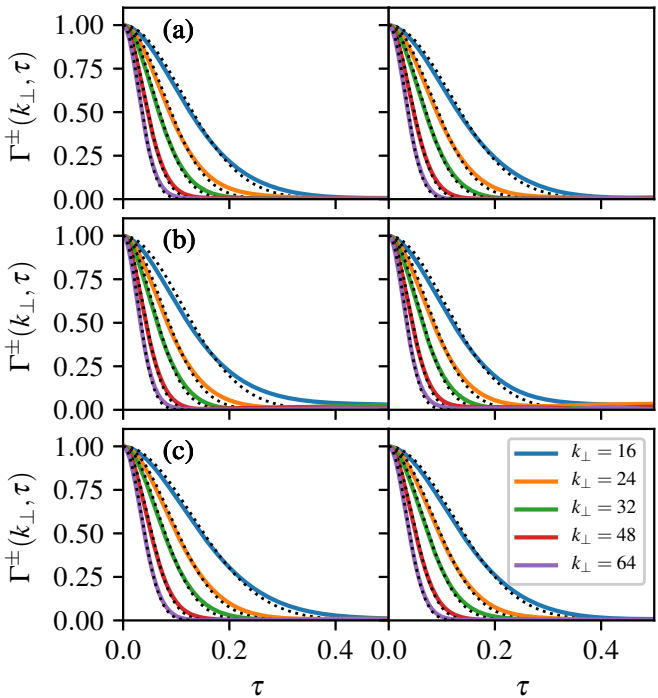}
  \caption{Scale-dependent time correlations $\Gamma^+(k_\perp,\tau)$ (left) and $\Gamma^-(k_\perp,\tau)$ (right) from simulations: (a) RB1, (b) RB2 and (c) RI1, for selected wavenumbers in the range $k_\perp=16$ to 64. Dotted lines correspond to Gaussian least-squares-fits with the decorrelation rate $\gamma_k$ are the only free parameter at each $k_\perp$. The inertial range of these simulations according to Figure~\ref{fig:h2d} is between $k_\perp\simeq 4$ and 20 for RB1 and RI1, and $k_\perp\simeq 4$ to 30 for RB2.}
  \label{fig:gammafigs}
\end{figure}

In the next section we present results from the scale-dependent time correlations measured in the simulations from equation~\eqref{eq:g2d}, and perform two important tests to compare with the theoretical formula  given in equations~\eqref{eq:two-time-spectrum-char} and \eqref{eq:two-time-spectrum} . The first test is to show that decorrelation rates $\gamma_k^\pm$ scale linearly with $k_\perp$, consistent with sweeping characterized by the \rms~speed of the outer-scale flow. In the second test the scale-dependent time correlations $\Gamma^\pm(k_\perp,\tau)$, as defined in equation~\eqref{eq:g2d}, are calculated in numerical simulations for wavenumbers $k_\perp$ in the inertial range. The resulting correlations are then compared with the characteristic function $\varphi_{\vec v'}(k_\perp\tau)$, computed from equation~\eqref{eq:pchar} using the random distribution of outer-scale velocities $P(\vec v')$. 

\begin{figure}[!t]
  \centering
  \includegraphics{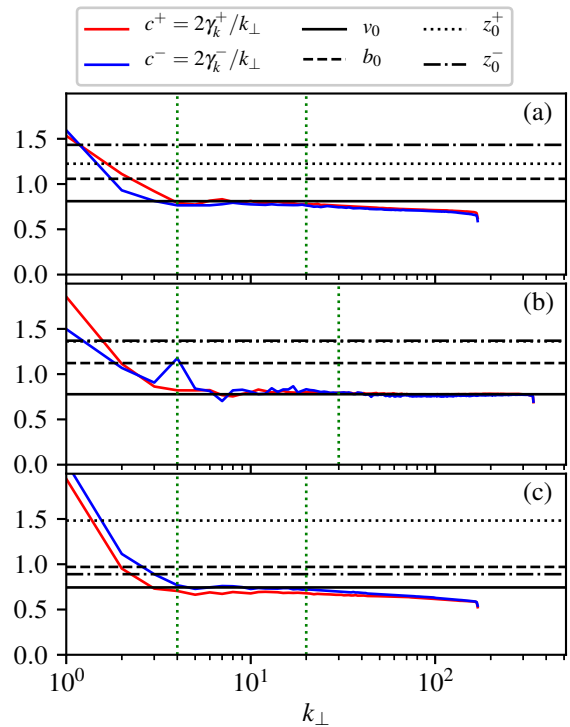}
  \caption{Sweeping velocities $c^\pm=2\gamma^\pm_k/k_\perp$ estimated from measured decorrelation rates $\gamma^\pm_k$ vs $k_\perp$, obtained from Gaussian fits shown in~Figure~\ref{fig:gammafigs}. The decorrelation rates are similar for both Elsasser fields at all wavenumbers $k_\perp$, for both balanced and imbalanced turbulence. Flat regions of $c^\pm$ correspond to wavenumbers for which the decorrelation is consistent with sweeping, and the characteristic sweeping speed agrees with \rms~of the velocity at the outer-scale. The \rms~values $z_0^\pm,b_0$ and $u_0$ at the outer scale, given in table~\ref{tab:sims}, are indicated as horizontal lines on the plot. The vertical dotted lines approximately represent the inertial range observed in the energy spectrum of each simulation in Figure~\ref{fig:h2d}.}
  \label{fig:gamma}
\end{figure}

 \section{Simulation Results.\label{sec:results}} 
 The modeled scale-dependent time correlations for strong MHD turbulence given in equation~\ref{eq:two-time-spectrum} have the following important features: 1) they are solely due to the random sweeping by the large-scale flow, 2) the decorrelation rates are the same for both Elsasser fluctuations $\vec z^\pm$, whether the turbulence is balanced or imbalanced, and scale linearly with $k_\perp$ as $\gamma^\pm=k_\perp u_0/2$ in the inertial range, and 3) they exhibit universal, self-similar behavior as they can all be written in terms of the characteristic function associated with the random distribution of large-scale velocities in the flow.

 The scale-dependent time correlations $\Gamma^\pm(k_\perp,\tau)$ are calculated for each simulation according to equation~\eqref{eq:g2d} and shown in Figure~\ref{fig:gammafigs} for selected values of the field-perpendicular wavenumber between $k_\perp=16$ and $k_\perp=64$. Dotted lines correspond to Gaussian fits of each correlation of the form given in Equation~\eqref{eq:two-time-spectrum} where the only free parameter is the corresponding decorrelation rate $\gamma_k^\pm$. As noted by~\citet{bourouaine18} in simulations of reflection-driven Alfv\'en turbulence, the scale-dependent time correlations closely follow Gaussian behavior in $\tau$, consistent with the statistics associated with the energy containing scales.

\begin{figure}[!t]
  \centering
  \includegraphics{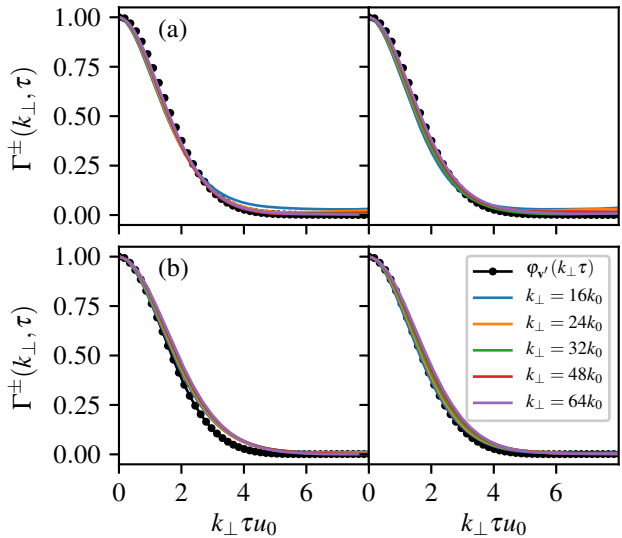}
  \caption{Scale-dependent time correlation functions of $\vec z^+$ (Left panels) and $\vec z^-$ (Right panels) vs $k_\perp\tau$ in simulations RB2 (a) and RI1 (b), for selected values of the wavenumber in the inertial range. The black line on all plots represent the characteristic function computed from the random distribution of velocities at the outer-scale. Similar results are found in balanced simulation~RB1. }
  \label{fig:two-time-power}
\end{figure}

Figure~\ref{fig:gamma} shows $c^\pm\equiv 2\gamma_k^\pm/k_\perp$ vs $k_\perp$ from the measured decorrelation rates for each Elsasser field $\vec z^\pm$, which is consistent with the prediction of BP19 in the strong turbulence regime  in a number of ways. First, the decorrelation rates for both fields remain approximately equal at all wavenumbers for both balanced and imbalanced simulations. Second, $c^\pm$ approximately starts to plateau at wavenumbers above $k_\perp\gtrsim 4$ and remains approximately constant up to wavenumbers around $k_\perp\simeq 20$ for simulations RB1 \& RI1 and up to $k_\perp\simeq 30$ for RB2, which indicates the decorrelation rates exhibit linear behavior consistent with sweeping in the inertial-range scales for each simulation. Moreover, the theoretical model predicts that $c^\pm$ is a common speed for both Elsasser variables $z^+$ and $z^-$ and equal to the characteristic sweeping speed $u_0$. Figure~\ref{fig:gamma} shows four horizontal lines corresponding to the \rms~values of $\vec v,\vec b,\vec z^\pm$ of the most energetic eddies, and $c^\pm$ is clearly consistent with the \rms~of the outer-scale velocities only, as indicated by the solid horizontal line.

The second test of the theoretical model is found in Figure~\ref{fig:two-time-power}, which shows the computed scale-dependent time correlation functions for selected values of $k_\perp$ in runs RB2 \& RI1 of balanced and imbalanced RMHD turbulence. In all these plots it is observed that when the corresponding temporal correlation function is plotted vs the normalized velocity-wavenumber $\hat\xi=k_\perp\tau u_0$, the scale dependent correlations for all wavenumbers are essentially indistinguishable from the characteristic function arising from the random distribution of velocities of the most energetic scales, as predicted by the model in equation~\eqref{eq:two-time-spectrum-char}.

\section{Conclusion.\label{sec:conclusion}} 

In this work we investigated the temporal decorrelation of strong MHD turbulence through phenomenological modeling and numerical simulations of RMHD turbulence. Scale-dependent time correlations of Elsasser fluctuations were modeled by restricting the recent BP19 phenomenology to the strong MHD turbulence regime. In the BP19 phenomenology, which extends Kraichnan's idealized convection model of Hydrodynamics to MHD, the Eulerian decorrelation is the result of HD sweeping and Alfv\'enic propagation along the local magnetic field. In this work we have shown that, for the particular case of strong turbulence regime, the decorrelation is solely dominated by HD sweeping. The resulting scale-dependent time correlations exhibit an universal, self-similar behavior that is entirely determined by the statistics of velocities at the largest, energy-containing eddies, and it clearly shows that the decorrelation rates for both Elssaser variables are the same regardless of whether the turbulence is balanced or imbalanced. All these features were tested using numerical simulations of strong RMHD turbulence. The numerical results are in very good agreement with the theoretical predictions.

An earlier model of the Eulerian decorrelation in MHD was also proposed by \citet{narita17a} in which the decorrelation rates are predicted to scale as $\gamma_k^\pm=k_\perp z^\mp_{\rm rms}/\sqrt{2}$. This  model is physically appealing at a first glance, as it is natural to assume that the decorrelation rate of ${\vec z}^+$ is determined by the \rms~value of ${\vec z'}^-$, and viceversa. However, as \cite{narita17a} points out, the decorrelations rates will be different for imbalanced turbulence where one of the Elsasser component has a larger amplitude than the other. One shortcoming of \citet{narita17a} model is that it does not take into account the fundamentally different effects that the large-scale flow and magnetic field fluctuations have on the small-scale ones. For instance, large-scale fluctuations in velocity will sweep small-scale eddies equally for both Elsasser fields, while large-scale fluctuations in the magnetic field simply modify the background magnetic field along which small-scale eddies propagate. It is therefore important that a sweeping model captures these different characteristics associated with the nature of large-scale velocity and magnetic field fluctuations.

The outcomes from this study will in fact be very beneficial for the analysis of the spacecraft signals beyond the validity of Taylor's Hypothesis whenever solar wind observations are compared with predictions from phenomenological MHD turbulence models~\citep[see for instance][]{bourouaine20}. Finally we conjecture that the results we present in this work may be suitably extended to kinetic (non-MHD) regimes found in the solar wind, but it requires further investigation.

\acknowledgments
This work was partially supported from NASA Living with a Start grant NNX16AH92G, Heliophysics Guest Investigator grant 80NSSC19K0275 and NSF-SHINE grant AGS-1752827. High-performance-computing resources were provided by the Argonne Leadership Computing Facility (ALCF) at Argonne National Laboratory, which is supported by the Office of Science of the U.S. Department of Energy under contract DE-AC02-06CH11357. The ALCF resources were granted under the INCITE program between 2012 and 2014. High-performance computing resources were also provided by the Texas Advanced Computing Center (TACC) at The University of Texas at Austin, under the NSF-XSEDE Project TG-ATM100031.
\end{document}